\documentclass[showpacs,prd]{revtex4}
\usepackage{epsfig}
\usepackage{graphicx}
\usepackage{dcolumn}
\usepackage{amsmath}
\usepackage{latexsym}

\begin{document} 
%%%%%%%%%%%%%%%%%%%%%%%%%%%%%%%%%%%%%%%%%%%%%%%%%%%%%%%%%%%%%%%%%%%%%%%%%%%%
%%%%%%%%%%%%%%%%%%%%%%%%%%%%%%%%%%%%%%%%%%%%%%%%%%%%%%%%%%%%%%%%%%%%%%%%%%%%

\begin{flushright}
NITheP-09-08\\
\end{flushright}

\title{Path integral action of a particle in the non commutative plane}  
\date{\today}
\author{Sunandan Gangopadhyay$^{a,c}$\footnote{e-mail: 
sunandan.gangopadhyay@gmail.com, sunandan@sun.ac.za}, 
Frederik G Scholtz $^{a,b}$\footnote{e-mail:
fgs@sun.ac.za}}
\affiliation{$^a$National Institute for Theoretical Physics (NITheP), 
Stellenbosch 7600, South Africa\\
$^b$Institute of Theoretical Physics, 
University of Stellenbosch, Stellenbosch 7600, South Africa\\
$^c$Department of Physics and Astrophysics, 
West Bengal State University, Barasat, India}

\begin{abstract}
\noindent Non commutative quantum mechanics can be viewed as a quantum system represented in the space of Hilbert-Schmidt operators acting on non commutative configuration space. Taking this as departure point, we formulate a coherent state approach to the path integral representation of the transition amplitude.  From this we derive an action for a particle moving in the non commutative plane and in the presence of an arbitrary potential.  We find that this action is non local in time.  However, this non locality can be removed by introducing an auxilary field, which leads to a second class constrained system that yields the non commutative Heisenberg algebra upon quantization. Using this action the propagator of the free particle and harmonic oscillator are computed explicitly.  For the harmonic oscillator this action yields the frequencies already obtained by other means in the literature.

\end{abstract}

\maketitle

%\noindent {\it{Introduction }}:\\
%\label{intro}

The idea of non commutative spacetime was first formally
introduced by Snyder in \cite{snyder} as an attempt to regulate the
divergences of quantum field theories. However, the discovery of renormalizable field theories pushed these ideas to the background until the difficulties encountered in the unification of gravity and quantum mechanics forced us to reconsider these ideas.  Indeed, strong arguments in favour of non commutative spacetime were given much more recently in \cite{dop} and further support for non commutative geometry came from string theory \cite{wit}. 

These observations led to a flurry of activities on non commutative quantum field theories \cite{doug} and the possible physical consequences of non commutative spacetime in quantum mechanics and quantum mechanical many-body systems \cite{duval}-\cite{khan}, quantum electrodynamics \cite{chai}-\cite{lia}, the standard model \cite{ohl} and cosmology  \cite{gar}, \cite{alex}. Despite this not much attention seems to have been paid to the formal and interpretational aspects of non commutative quantum mechanics and, particularly, the path integral formulation and derivation of the action for a particle moving in a non commutative space. Indeed, some of these results are even in disagreement. 

Non commutative geometry implies the absence of
common position eigenkets. This problem was  circumvented in \cite{spallucci}
 by taking coherent states to define the 
propagation kernel. The coherent states, being the eigenstates
of complex combinations of the position operators, act as a
meaningful replacement for the position eigenstates admissible
only in the commutative theory. The observation 
made in this paper is that the free particle propagator turns
out to exhibit an ultra-violet cutoff induced by the non commutative
parameter $\theta$. However, the non commutative version of the path integral
derived in this paper does not reflect the expected breaking of time reversal in non commutative systems \cite{muthukumar, scholtz, laure}.  In another paper \cite{tan}, the same trick
of using coherent states to define the propagation kernel
was employed, however, the final expressions for the propagator
and the resulting physics were quite different from \cite{spallucci}.  Closer analysis reveals that the limit of infinite time slices is rather subtle and has to be done with great care, which is one of the points emphasized in this paper.

Recently \cite{laure} suggested that non commutative quantum mechanics should be formulated in the Hilbert space of Hilbert-Schmidt operators acting on classical configuration space. Within this context a positive operator valued measure (POVM) for position measurement was constructed as coherent states on the Hilbert space of Hilbert-Schmidt operators. Apart from providing a consistent prescription for position measurement on a non commutative space, these coherent states also provide a resolution of the identity, which quite naturally introduces the Voros product on the Hilbert space of Hilbert-Schmidt operators.  As such these states can be used to derive a path integral representation for the transition amplitude and, particularly, the action for a particle moving in a non commutative plane, which is the theme of this paper.  Not surprisingly, the resulting action turns out to be non local in time.  However, we also show that this non locality can be avoided in favour of a second class constraint system, which yields upon quantization the non commutative Heisenberg algebra.  We proceed to use the non local action to compute the propagator for the free particle and the harmonic oscillator.  As a by product, we show that the equation of motion for the harmonic oscillator already yields the correct frequencies of the non commutative harmonic oscillator as obtained through a Bopp shift. 

We start with a brief review of the key elements in the above construction that we need to derive the path integral representation of the transition amplitude.  In two dimensions, the coordinates of non commutative configuration space satisfy the commutation relation 
\begin{equation}
[\hat{x}, \hat{y}] = i\theta
\label{1}
\end{equation} 
where, without loss of generality, it is assumed that $\theta$ is a real positive parameter. Defining annihilation and creation operators
$b = \frac{1}{\sqrt{2\theta}} (\hat{x}+i\hat{y})$,
$b^\dagger =\frac{1}{\sqrt{2\theta}} (\hat{x}-i\hat{y})$
that satisfy the Fock algebra $[ b ,b^\dagger ] = 1$, non commutative configuration space is isomorphic to boson Fock space
\begin{eqnarray}
\mathcal{H}_c = \textrm{span}\{ |n\rangle= 
\frac{1}{\sqrt{n!}}(b^\dagger)^n |0\rangle\}_{n=0}^{n=\infty}
\label{3}
\end{eqnarray}
where the span is take over the field of complex numbers.

The Hilbert space of the non commutative quantum system is the set of Hilbert-Schmidt operators acting on non commutative configuration space
\begin{equation}
\mathcal{H}_q = \left\{ \psi(\hat{x},\hat{y}): 
\psi(\hat{x},\hat{y})\in \mathcal{B}
\left(\mathcal{H}_c\right),\;
{\rm tr_c}(\psi^\dagger(\hat{x},\hat{y})
\psi(\hat{x},\hat{y})) < \infty \right\}.
\label{4}
\end{equation}
Here ${\rm tr_c}$ denotes the trace over non commutative 
configuration space and $\mathcal{B}\left(\mathcal{H}_c\right)$ 
the set of bounded operators on $\mathcal{H}_c$.  When equipped with the trace inner product, this space forms a Hilbert space \cite{hol}.  To distinguish states in the non commutative configuration space (\ref{3}) from those in the quantum Hilbert space (\ref{4}), we denote states in the former by $|\cdot\rangle$ and states in the latter by $\psi(\hat{x},\hat{y})\equiv |\psi)$. 

Assuming commutative momenta (this can be generalized as discussed in \cite{laure}) the abstract non commutative Heisenberg algebra in two dimensions reads 
\begin{eqnarray}
\label{hei}
\left[\hat{x},\hat{p}_x\right]&=& i\hbar\quad,
\quad\left[\hat{y},\hat{p}_y\right]= i\hbar\nonumber\\
\left[\hat{x}, \hat{y}\right]&=&i\theta\quad,\quad
\left[\hat{p}_x,\hat{p}_y\right]= 0.
\end{eqnarray}
A unitary representation of this algebra in terms of operators 
$\hat{X}$, $\hat{Y}$, $\hat{P}_x$ and $\hat{P}_y$ 
acting on the quantum Hilbert space (\ref{4}) is easily found to be 
\begin{eqnarray}
\label{schnc}
\hat{X}\psi(\hat{x},\hat{y}) &=& \hat{x}\psi(\hat{x},\hat{y})\quad,\quad
\hat{Y}\psi(\hat{x},\hat{y}) = \hat{y}\psi(\hat{x},\hat{y})\nonumber\\
\hat{P}_x\psi(\hat{x},\hat{y}) &=& \frac{\hbar}{\theta}[\hat{y},\psi(\hat{x},\hat{y})]\quad,\quad
\hat{P}_y\psi(\hat{x},\hat{y}) = -\frac{\hbar}{\theta}[\hat{x},\psi(\hat{x},\hat{y})]~.
\label{action}
\end{eqnarray}
Note that the position acts by left multiplication and the momentum adjointly.  We use capital letters to distinguish operators acting on quantum Hilbert space from those acting on non commutative configuration space. It is also useful to introduce the following operators on the quantum Hilbert space 
$B=\frac{1}{\sqrt{2\theta}}\left(\hat{X}+i\hat{Y}\right)$,
$B^\dagger=\frac{1}{\sqrt{2\theta}}\left(\hat{X}-i\hat{Y}\right)$,
$\hat{P}=\hat{P}_x + i\hat{P}_y$ and
$\hat{P}^\dagger = \hat{P}_x -i\hat{P}_y$.
We note that $\hat{P}^2=\hat{P}^2_x+\hat{P}^2_y 
= P^\dagger P = PP^\dagger$. These operators act in the following way
\begin{eqnarray}
\label{action1}
B\psi(\hat{x},\hat{y}) &=& b\psi(\hat{x},\hat{y})\quad,\quad
B^\dagger\psi(\hat{x},\hat{y}) = b^\dagger\psi(\hat{x},\hat{y})\nonumber\\
P\psi(\hat{x},\hat{y})&=& -i\hbar \sqrt{\frac{2}{\theta}}[b,\psi(\hat{x},\hat{y})]
\quad,\quad P^\dagger\psi(\hat{x},\hat{y}) = i\hbar
\sqrt{\frac{2}{\theta}}[ b^{\dagger},\psi(\hat{x},\hat{y})].
\end{eqnarray}

The minimal uncertainty states on non commutative 
configuration space are well known to be the normalized coherent states \cite{klaud}
\begin{equation}
\label{cs} 
|z\rangle = e^{-z\bar{z}/2}e^{z b^{\dagger}} |0\rangle
\end{equation}
where, $z=\frac{1}{\sqrt{2\theta}}\left(x+iy\right)$ 
is a dimensionless complex number. These states provide an overcomplete 
basis on the non commutative configuration space. 
Corresponding to these states we can construct a state 
(operator) in quantum Hilbert space as follows
\begin{equation}
|z, \bar{z} )=\frac{1}{\sqrt{2\pi\theta}}|z\rangle\langle z|.
\label{csqh}
\end{equation}
The particular choice of normalization will be explained in a moment. It is easily checked that these operators are indeed of Hilbert-Schmidt type and that they also satisfy
\begin{equation}
B|z, \bar{z})=z|z, \bar{z}).
\label{p1}
\end{equation}

To construct the path integral we also need to introduce momentum eigenstates
\begin{eqnarray}
|p)=\sqrt{\frac{\theta}{2\pi\hbar^{2}}}e^{i\sqrt{\frac{\theta}{2\hbar^2}}
(\bar{p}b+pb^\dagger)}
\label{eg}
\end{eqnarray}
satisfying
\begin{eqnarray}
\hat{P}_{x}|p)=p_{x}|p)\quad,\quad\hat{P}_{y}|p)=p_{y}|p)
\label{eg1}
\end{eqnarray}
\begin{eqnarray}
(p'|p)=e^{-\frac{\theta}{4\hbar^{2}}(\bar{p}p+\bar{p}'p')}
e^{\frac{\theta}{2\hbar^{2}}\bar{p}p'}\delta(p-p')~.
\label{eg2}
\end{eqnarray}

Crucial to the path integral construction are the following completeness relations \cite{laure} 
\begin{eqnarray}
\int d^{2}p~|p)(p|=1_{Q}~,
\label{eg5}
\end{eqnarray}
and
\begin{eqnarray}
\int 2\theta dzd\bar{z}~|z, \bar{z})\star(z, \bar{z}|=\int dxdy~|z, \bar{z})\star(z, \bar{z}|=1_{Q},
\label{eg6}
\end{eqnarray}
where the star product between two functions 
$f(z, \bar{z})$ and $g(z, \bar{z})$ is defined as
\begin{eqnarray}
f(z, \bar{z})\star g(z, \bar{z})=f(z, \bar{z})
e^{\stackrel{\leftarrow}{\partial_{\bar{z}}}
\stackrel{\rightarrow}{\partial_z}} g(z, \bar{z})~.
\label{eg7}
\end{eqnarray}
The choice of normalization in (\ref{csqh}) is dictated by the requirement that the completeness relation, when written in terms of the dimensionfull coordinates $x$,$y$, yields the same measure as in the commutative case, i.e., $dxdy$ as displayed in the second equality of eq. (\ref{eg6}).     
   
The following overlap also plays an important role in the derivation
\begin{eqnarray}
(z, \bar{z}|p)=\frac{1}{\sqrt{2\pi\hbar^{2}}}
e^{-\frac{\theta}{4\hbar^{2}}\bar{p}p}
e^{i\sqrt{\frac{\theta}{2\hbar^{2}}}(p\bar{z}+\bar{p}z)}~.
\label{eg3}
\end{eqnarray}

With these results and completeness relations for the
momentum and the position eigenstates 
(\ref{eg5}, \ref{eg6}) in place, we can proceed to write down the
path integral representation of the propagation kernel
in two dimensional non commutative space. This reads
\begin{eqnarray}
(z_f, t_f|z_0, t_0)&=&\lim_{n\rightarrow\infty}\int \left(2\theta\right)^{n}
(\prod_{j=1}^{n}dz_{j}d\bar{z}_{j})~(z_f, t_f|z_n, t_n)\star_n
(z_n, t_n|....|z_1, t_1)\star_1(z_1, t_1|z_0, t_0)\nonumber\\
&=&\lim_{n\rightarrow\infty}\int_{-\infty}^{+\infty}
(\prod_{j=1}^{n}dx_{j}dy_{j})~(z_f, t_f|z_n, t_n)\star_n
(z_n, t_n|....|z_1, t_1)\star_1(z_1, t_1|z_0, t_0)~.
\label{pint1}
\end{eqnarray}
The next step is to compute the propagator over a small segment in the
above path integral. With the help of (\ref{eg3}) and (\ref{eg5}),
we have
\begin{eqnarray}
(z_{j+1}, t_{j+1}|z_j, t_j)&=&(z_{j+1}|e^{-\frac{i}{\hbar}H\tau}|z_j)\nonumber\\
&=&(z_{j+1}|1-\frac{i}{\hbar}H\tau +O(\tau^2)|z_j)\nonumber\\
&=&\int_{-\infty}^{+\infty}d^{2}p_j~e^{-\frac{\theta}{2\hbar^{2}}p_{j}^{2}}
e^{i\sqrt{\frac{\theta}{2\hbar^{2}}}\left[p_{j}(\bar{z}_{j+1}-\bar{z}_{j})+\bar{p}_{j}(z_{j+1}-z_{j})\right]}e^{-\frac{i}{\hbar}\tau[\frac{p_{j}^2}{2m}+V(\bar{z}_{j+1}, z_{j})]}+O(\tau^2)
\label{pint2}
\end{eqnarray} 
where, $H=\frac{\vec{P_i}^2}{2m}+:V(B^{\dagger},B):$ is the Hamiltonian acting on the quantum Hilbert space and $V(\hat{X},\hat{Y})$ is the normal ordered potential expressed
in terms of the annihilation and creation operators ($B$, $B^{\dagger}$). 
Substituting the above expression in (\ref{pint1}) and computing the star products explicitly, we obtain (apart from a constant factor)
\begin{eqnarray}
(z_f, t_f|z_0, t_0)=&&\int\prod_{j=1}^{n}(dx_{j}dy_{j})
\prod_{j=0}^{n}d^{2}p_{j}
\exp\left(\sum_{j=0}^{n}\left[i\sqrt{\frac{\theta}{2\hbar^{2}}}\left[p_{j}(\bar{z}_{j+1}-\bar{z}_{j})+\bar{p}_{j}(z_{j+1}-z_{j})\right]+\alpha p_{j}\bar{p}_{j}
-\frac{i}{\hbar}\tau V(\bar{z}_{j+1},z_{j})\right]\right.\nonumber\\
&&\left.~~~~~~~~~~~~~~~~~~~~~~~~~~~~~~~~~~~~~~~~~~~~~~~~~~~+\frac{\theta}{2\hbar^{2}}\sum_{j=0}^{n-1}p_{j+1}\bar{p}_{j}\right)
\label{pint3}
\end{eqnarray} 
where $\alpha=-(\frac{i\tau}{2m\hbar}+\frac{\theta}{2\hbar^{2}})$.  An important difference between the commutative and non commutative cases is that in the latter the momentum integral involves off diagonal terms that couple $p_j$ and $p_{j+1}$.  This already suggests a possible non locality in time once the momentum integral, which is now considerably more complicated, has been performed. To perform the momentum integral it is convenient to make the following identification $p_{n+1}=p_{0}$ and to recast the integrand of the above integral into the following form:
$$
\exp\left(-\vec{\partial}_{z_{f}}\vec{\partial}_{\bar{z}_{0}}\right)\exp\left(\sum_{j=0}^{n}\left[i\sqrt{\frac{\theta}{2\hbar^{2}}}\left[p_{j}(\bar{z}_{j+1}-\bar{z}_{j})+\bar{p}_{j}(z_{j+1}-z_{j})\right]+\alpha p_{j}\bar{p}_{j}-\frac{i}{\hbar}\tau V(\bar{z}_{j+1},z_{j})+\frac{\theta}{2\hbar^{2}}p_{j+1}\bar{p}_{j}\right]
\right).
$$
The purpose of the boundary operator in the above expression is to cancel an additional coupling that has been introduced between $p_0$ and $p_n$.  The reason for introducing this coupling is that the resulting matrix that needs to be inverted when performing the momentum integral is now of a simple invertible form.  Indeed, the momentum integral is of Gaussian form
\begin{equation}
\prod_{j=0}^{n}d^{2}p_{j}
\exp\left(\sum_{i,j}p_{i}A_{i,j}\bar{p}_{j}\right),
\label{pint4}
\end{equation} 
where $A$ is a $D\times D$ dimensional matrix with $D=n+1=T/\tau$, $T=t_{f}-t_{0}$ and the identification $n+1\equiv 0$ is made, given by
\begin{equation}
A_{lr}=\alpha\delta_{l,r}+\frac{\theta}{2\hbar^{2}}\delta_{l+1,r}~.
\label{matrix}
\end{equation}
A simple inspection shows that the eigenvalues and ortho-normal eigenvectors of the
matrix $A$ are given by
\begin{eqnarray}
\lambda_{k}&=&\alpha+\frac{\theta}{2\hbar^{2}}e^{2\pi ik/D}\quad;\quad k\in[0,n]\nonumber\\
u_{k}&=&\frac{1}{\sqrt{D}}(1\quad e^{2\pi ik/D}\quad e^{4\pi ik/D}....)^{T}.
\label{evalues}
\end{eqnarray} 
Since the real part of the eigenvalues of $A$ are non positive, the momentum integral can be performed to obtain
\begin{eqnarray}
(z_f, t_f|z_0, t_0)=N\int\prod_{j=1}^{n}(dz_{j}d\bar{z}_{j})
\exp\left(-\vec{\partial}_{z_{f}}\vec{\partial}_{\bar{z}_{0}}\right)
\exp\left(\frac{\theta}{2\hbar^{2}}\sum_{l=0}^{n}\sum_{r=0}^{n}
(\bar{z}_{l+1}-\bar{z}_{l})A^{-1}_{lr}(z_{r+1}-z_{r})\right)
\exp\left(-\frac{i}{\hbar}\tau \sum_{j=0}^{n}V(\bar{z}_{j+1},z_{j})\right).\nonumber\\
\label{pintegral1}
\end{eqnarray} 
The inverse of the matrix $A$ is easily obtained as $A^{-1}_{lr}=\frac{1}{D}\sum_{k=0}^{n}\lambda_{k}^{-1}e^{-2\pi i(l-r)k/D}$ which yields
\begin{eqnarray}
(z_f, t_f|z_0, t_0)=&&N\int\prod_{j=1}^{n}(dz_{j}d\bar{z}_{j})
\exp\left(-\vec{\partial}_{z_{f}}\vec{\partial}_{\bar{z}_{0}}\right)
\exp\left(\frac{\theta\tau}{2\hbar^{2}T}\sum_{l,r,k=0}^{n}
\tau\dot{\bar{z}}(l\tau)[\alpha+\frac{\theta}{2\hbar^{2}}e^{-\tau\partial_{(l\tau)}}]^{-1}
[e^{-2\pi i(l-r)k\tau/T}]\tau\dot{z}(r\tau)\right)\nonumber\\
&&~~~~~~~~~~~~~~~~~~~~~~~~~\times\exp\left(-\frac{i}{\hbar}\tau \sum_{j=0}^{n}V(\bar{z}_{j},z_{j})+O(\tau^{2})\right)
\nonumber\\
=&&N\int\prod_{j=1}^{n}(dz_{j}d\bar{z}_{j})\exp\left(-\vec{\partial}_{z_{f}}\vec{\partial}_{\bar{z}_{0}}\right)\nonumber\\
&&~~~~~~~~~~~~~~~~~~~~\times\exp\left(\frac{\theta}{2\hbar^{2}T}\sum_{l,r,k=0}^{n}
\tau\dot{\bar{z}}(l\tau)[-\frac{i}{2m\hbar}-\frac{\theta}{2\hbar^{2}}\partial_{(l\tau)}+O(\tau)]^{-1}[e^{-2\pi i(l-r)k\tau/T}]\tau\dot{z}(r\tau)\right)\nonumber\\
&&~~~~~~~~~~~~~~~~~~~~\times\exp\left(-\frac{i}{\hbar}\tau \sum_{j=0}^{n}V(\bar{z}_{j},z_{j})+O(\tau^{2})\right).
\label{pintegral2}
\end{eqnarray} 
In the second line we have used the fact that $z_{l}=z(l\tau)$
and $z_{l+1}-z_{l}=\tau\dot{z}(l\tau)+O(\tau^{2})$. Performing the sum over $k$ and taking the 
limit $\tau\rightarrow 0$, we finally find
the path integral representation of the propagator
\begin{eqnarray}
(z_f, t_f|z_0, t_0)&=&N\int \mathcal{D}z\mathcal{D}\bar{z}
\exp\left(-\vec{\partial}_{z_{f}}\vec{\partial}_{\bar{z}_{0}}\right)
\exp({\frac{i}{\hbar}S})
\label{pintegral3}
\end{eqnarray} 
where $S$ is the action given by
\begin{eqnarray}
S=\int_{t_{0}}^{t_{f}}dt~ \left[\frac{\theta}{2}\dot{\bar{z}}(t)(\frac{1}{2m}-\frac{i\theta}{2\hbar}
\partial_{t})^{-1}
\dot{z}(t)- V(\bar{z}(t),z(t))\right]~.
\label{action_ncqm}
\end{eqnarray}
Note that the first order derivative in time gives the expected time reversal symmetry breaking.

As a consistency check, we quantize this theory to see if we recover the non commutative Heisenberg algebra (\ref{hei}). To do this we first introduce an auxilary complex field $\phi$ and write the transition amplitude as
\begin{equation}
(z_f, t_f|z_0, t_0)=\tilde N\int \mathcal{D}\phi\mathcal{D}\bar{\phi}\int \mathcal{D}z\mathcal{D}\bar{z}
\exp\left(-\vec{\partial}_{z_{f}}\vec{\partial}_{\bar{z}_{0}}\right)
\exp({\frac{i}{\hbar}S})~,
\label{pintegral4}
\end{equation} 
with action
\begin{equation}
\label{auxact}
S=\int_{t_{0}}^{t_{f}}dt~\left[\bar\phi\left(\frac{i}{\hbar}\partial_t-m\theta\right)\phi+\bar\phi\dot z+\phi\dot{\bar z}-V\left(\bar z,z\right)\right].
\end{equation}
Writing $\phi=\phi_1+i\phi_2$, $z=\frac{1}{\sqrt{2\theta}}\left(x+iy\right)$ one finds that this is a constrained system with constraints $\chi_1=p_{\phi_1}-\frac{1}{\hbar}\phi_2$, $\chi_2=p_{\phi_2}+\frac{1}{\hbar}\phi_1$, $\chi_3=p_{x}-\sqrt{\frac{2}{\theta}}\phi_1$ and $\chi_4=p_{y}-\sqrt{\frac{2}{\theta}}\phi_2$.  Computing the Poisson brackets shows that these constraints are second class.  Introducing the Dirac bracket and replacing $\{\cdot,\cdot\}_{DB}\rightarrow \frac{1}{i\hbar}[\cdot,\cdot]$, indeed yields the non commutative Heisenberg algebra (\ref{hei}).  Furthermore all the Dirac brackets of the constraints with the Hamiltonian vanish.

We now compute the propagator for the free particle from the above path integral
representation of the propagator. The classical equation of motion is easily obtained from the above action as
\begin{eqnarray}
K\ddot{z}_{c}(t)=0\quad;\quad K=(\frac{1}{2m}-\frac{i\theta}{2\hbar}\partial_{t})^{-1}~.
\label{eom}
\end{eqnarray} 
Transforming to Fourier space, one easily verifies that the operator $K$ has a trivial kernel, which leads to the same equation of motion as in the commutative case.  Solving this equation subject to the boundary conditions that at 
$t=t_{0}$, $z_{c}=z_{0}$, $t=t_{f}$, $z_{c}=z_{f}$, we obtain
\begin{eqnarray}
z_{c}(t)=z_{0}+\frac{z_{f}-z_{0}}{T}(t-t_{0})~.
\label{solution}
\end{eqnarray} 
Substituting the above solution in (\ref{pintegral3}) (with $V=0$) and acting with the boundary operator, we obtain
\begin{eqnarray}
(z_f, t_f|z_0, t_0)&=&N
\exp{\left[-\frac{m}{2(i\hbar T+m\theta)}
(\vec{x}_{f}-\vec{x}_{0})^2\right]}\quad;\quad (n+1)\tau=T=t_{f}-t_{0}
\label{free_prop}
\end{eqnarray}
where we have included the contribution coming from the fluctuations in the normalisation constant $N$. The easiest way to determine this constant is to use the following identity
\begin{eqnarray}
(z_f, t_f|p)&=&2\theta\int~dz_{0}d\bar{z}_{0}~
(z_f, t_f|z_0, t_0)\star_{0}(z_0, t_0|p)~.
\label{identity}
\end{eqnarray}
Computing the right hand side using (\ref{eg3}) and comparing with the left hand side fixes
the constant $N=\frac{m}{2\pi\left(\theta m+i\hbar T\right)}$. 

We now include the harmonic oscillator potential 
$V=\frac{1}{2}m\omega^2(\hat{X}^{2}+\hat{Y}^{2})$ in the
Hamiltonian. Up to a constant the normal ordered form of this potential
in terms of the creation and annihilation operators is $:V:=m\omega^2\theta B^{\dagger}B$.
The action for the harmonic oscillator therefore reads
\begin{eqnarray}
S=\int_{t_{0}}^{t_{f}}dt~ \theta\left[\frac{1}{2}\dot{\bar{z}}(t)(\frac{1}{2m}-\frac{i\theta}{2\hbar}\partial_{t})^{-1}
\dot{z}(t)-m\omega^2 \bar{z}(t)z(t)\right]~.
\label{action_nchar}
\end{eqnarray} 
The equation of motion following from the above action is
\begin{eqnarray}
K\ddot{z}_{c}(t)+2m\omega^2 z_{c}(t)&=&0~.
\label{harsol}
\end{eqnarray} 
Making the following ansatz for the solution
\begin{eqnarray}
z_{c}(t)&=&e^{i\gamma t},
\label{ansatz}
\end{eqnarray} 
yields the frequencies 
\begin{eqnarray}
\gamma_\pm&=&\frac{1}{2\hbar}(m\omega^{2}\theta\pm\omega\sqrt{m^{2}\omega^{2}\theta^{2}+4\hbar^{2}}).
\label{energyeig}
\end{eqnarray} 
Noting that the one frequency is positive and the other negative, we can write the general classical solution as 
\begin{equation}
z_c(t)=a_+e^{i\omega_+ t}+a_-e^{-i\omega_- t}
\label{gensol}
\end{equation}
where we have introduced the positive frequencies $\omega_+=\gamma_+$ and $\omega_-=-\gamma_-$.  This already reflects a breaking of time reversal symmetry as these frequencies are not identical. Note, however, that they coincide when $\theta=0$.  As already remarked earlier time reversal symmetry breaking is a generic feature of the action (\ref{action_ncqm}) arising from the first order derivative in time.  Remarkably these are exactly the frequencies that arise when the non commutative harmonic oscillator is solved by other means \cite{muthukumar,laure}.

We can now proceed to compute the harmonic oscillator propagator.  From the boundary conditions that at 
$t=t_{0}$, $z_{c}=z_{0}$, $t=t_{f}$, $z_{c}=z_{f}$, one can solve the coefficients $a_\pm$ appearing in (\ref{gensol}) in terms of $z_0$ and $z_f$.  Substituting (\ref{gensol}) into the action, absorbing the fluctuations and all other constants in the normalization and acting with the boundary operator yields
\begin{eqnarray}
(z_f, t_f|z_0, t_0)&=&N\exp\left(Z^\dagger\Lambda Z\right)
\label{harosprop}
\end{eqnarray}
where $Z^\dagger=\left(z_0^*,z_f^*\right)$ and $\Lambda$ is the $2\times 2$ matrix
\begin{eqnarray}
\Lambda = \frac{m\theta}{m\theta Q_{12}-\hbar}\left(
\begin{array}{cc}Q_{11} & Q_{12} \\
Q_{21}+\frac{m\theta\omega_+\omega_-}{\hbar}& Q_{22}
\end{array}\right).
\end{eqnarray}
with $Q$ the matrix
\begin{eqnarray}
Q = \left(
\begin{array}{cc}
-\frac{\omega_-+\omega_+e^{iT(\omega_-+\omega_+)}}{1-e^{iT(\omega_-+\omega_+)}} & \frac{\omega_-+\omega_+}{e^{-iT\omega_+}-e^{iT\omega_-}} \\
\frac{\omega_-+\omega_+}{e^{-iT\omega_-}-e^{iT\omega_+}}& -\frac{\omega_++\omega_-e^{iT(\omega_-+\omega_+)}}{1-e^{iT(\omega_-+\omega_+)}}
\end{array}\right).
\end{eqnarray}

The normalization can be fixed in the same way as for the free particle.  Here the simplest is to study the time evolution of the harmonic oscillator ground state, which has been computed in \cite{laure}.  The result is
\begin{equation}
N=\frac{1}{2\pi}\left[\frac{m\omega_-}{\hbar}-\left(1+\frac{m\omega_-\theta}{\hbar}\right)\frac{mQ_{11}}{m\theta Q_{12}-\hbar}\right]e^{-\frac{iT}{2}\left(\omega_-+\omega_+\right)}
\end{equation} 

Taking the limit $\theta\rightarrow 0$ yields for the free particle and the harmonic oscillator the precise commutative propagator.

In this paper we have formulated the path integral
representation of the propagation kernel on the non commutative plane using the recently proposed formulation of non commutative quantum mechanics as a quantum system on Hilbert-Schmidt space. From the path integral, we have obtained the action of a particle moving in the non commutative plane and an arbitrary potential.  This is the central result of this paper. This action is non local in time, but this non locality can be removed by introducing an auxilary field, which leads to a second class constrained system that yields the non commutative Heisenberg algebra upon quantization. We solved the equation of motion for a free particle and using this we explicitly computed the path integral and thus propagator of the free particle.  This propagator exhibits a ultra-violet
cutoff induced by the non commutative parameter. We also solved the equation of motion for the harmonic oscillator and found the result to be in conformity with existing results in the literature obtained by other methods.  In particular explicit breaking of time reversal symmetry was found.  We also computed the propagator of the harmonic oscillator.  For both the free particle and harmonic oscillator the propagators reduce to the commutative result in the limit $\theta\rightarrow 0$. 

%%%%%%%%%%%%%%%%%%%%%%%%%%%%%%%%%%%%%%%%%%%%%%%%%%%%%%%%%%%%
\noindent {\bf{Acknowledgements}} : This work was supported under a grant of the National Research Foundation of South Africa. 

%%%%%%%%%%%%%%%%%%%%%%%%%%%%%%%%%%%%%%%%%%%%%%%%%%%%%%%%%%%%%%%%%%%%%%%%%%%

%%%%%%%%%%%%%%%%%%%%%%%%%%%%%%%%%%%%%%%%%%%%%%%%%%%%%%%%%%%%%%%%%%%%%%%%%%%%
%%%%%%%%%%%%%%%%%%%%%%%%%%%%%%%%%%%%%%%%%%%%%%%%%%%%%%%%%%%%%%%%%%%%%%%%%%%%
\end{document}